\documentclass{article}
\usepackage[utf8]{inputenc}
\usepackage{graphicx}
\usepackage{hyperref}
\usepackage{color}

\title{Towards Understanding Provenance in Industry}
\date{}
\author{Matthias Galster\footnote{University of Canterbury, New Zealand, matthias.galster@canterbury.ac.nz} \and Jens Dietrich\footnote{Victoria University of Wellington, New Zealand, jens.dietrich@vuw.ac.nz}}

\begin{document}

\maketitle

\section*{Abstract}

\textbf{Context:} Trustworthiness of software has become a first-class concern of users (e.g., to understand software-made decisions). Also, there is increasing demand to demonstrate regulatory compliance of software and end users want to understand how software-intensive systems make decisions that affect them.

\noindent \textbf{Objective:} We aim to provide a step towards understanding provenance needs of the software industry to support trustworthy software. Provenance is information about entities, activities, and people involved in producing data, software, or output of software, and used to assess software quality, reliability and trustworthiness of digital products and services.

\noindent \textbf{Method:} Based on data from in-person and questionnaire-based interviews with professionals in leadership roles we develop an ``influence map'' to analyze \emph{who} drives provenance, \emph{when} provenance is relevant, \emph{what} is impacted by provenance and \emph{how} provenance can be managed. 

\noindent \textbf{Results:} The influence map helps decision makers navigate concerns related to provenance. It can also act as a checklist for initial provenance analyses of systems. It is empirically-grounded and designed bottom-up (based on perceptions of practitioners) rather than top-down (from regulations or policies).

\noindent \textbf{Conclusion:} We present an imperfect first step towards understanding provenance based on current perceptions and offer a preliminary view ahead.

\section{Introduction}
\label{sec:introduction}

Software is often built from components. Components on the other hand are designed as \emph{black boxes}, 
 offering functionality through an interface and hiding implementations. When using components, the focus is typically on \emph{what} functionality they offer. However, there is growing interest in \emph{how} functionality is achieved~\cite{chazette2020explainabilitynfr}. This is mostly due to two trends:  (1) Software-supported decisions using AI or ML capabilities led to components that are even ``more black box'' that not even  engineers understand~\cite{defranco2022algospuppeteers}. This raises the question of how to trust systems composed of ``black boxes''. (2) New  regulatory requirements demand approaches to provide evidence for compliance of software (AI-supported or not), potentially for different stakeholders~\cite{chauhan2022AIdei}, for example, regulators in the finance sector~\cite{daniel2009business}. Similarly, users may worry about transparency of software-supported decisions~\cite{chazette2020explainabilitynfr}. The question arises how data used in software is being collected and used to make decisions, and whether software  disadvantages users. For instance, it has been demonstrated that Amazon's recruitment software is biased against women~\footnote{https://www.reuters.com/article/us-amazon-com-jobs-automation-insight-idUSKCN1MK08G [last access: September 9, 2022]}. Job applicants wanting to contest recruitment decisions need information about data and algorithms used to derive decisions. 
This also links to ethical considerations in technology~\cite{Vakkuri2020AIethics}.

In this short communication, we provide one step towards understanding the needs in industry regarding \emph{provenance} and how to support trustworthy software. Provenance is ``metadata about the origin, context or history of data~\cite{cheney2009provenance}.'' In our work, provenance goes beyond data and considers ``information about entities, activities, and people involved in producing  data or software, or output of software, which can be used assess its quality, reliability or trustworthiness.'' A pre-requisite for reliability and trustworthiness is interpretability~\cite{lipton2018mythos}. Interpretability is about transparency 
and explainability 
~\cite{lipton2018mythos}. Therefore, provenance eventually enables trust, transparency and explainability. 

We provide the following \textbf{contribution}: From 24 questionnaire-based and in-person interviews with professionals in technical leadership positions we develop an ``influence map'' to analyze provenance in terms of stakeholders, drivers, requirements, and practices. This influence map can also be a checklist for initial provenance analyses of systems. It is empirically-grounded and designed bottom-up (perceptions of practitioners) rather than top-down starting from regulations or policies.

\section{Method}
\label{sec:methhod}

We conducted seven in-person interviews with individuals in technical leadership roles (CTO, CEO, lead architects, etc.) from different companies and domains. At the same time, we collected data via 17 questionnaire-based interviews with individuals answering a questionnaire in their own time. This was suitable for individuals in leadership positions with limited time. Also, we completed this work during COVID, so were limited when  conducting in-person interviews. The questionnaire-based interviews were not to conduct a large-scale survey, but to offer interviewees an alternative. Therefore, the questionnaire was a guided form  of the in-person interviews. 

Companies of interviewees are medium and large scale, globally operating and exporting, their products include AI and ML capabilities, and they offer APIs to external systems and consume APIs. Companies are from ``veracity-enabling'' domains (i.e., providers of technology to support trustworthiness and explainability), e.g., cloud engineering, testing, application monitoring, but also from ``veracity-intensive'' domains (i.e., adopters of veracity technology), e.g., financial services, Health IT.

We analyzed data using an inductive, data-driven approach. We did not have a pre-defined coding protocol and did not pre-define  drivers, stakeholders, etc. of provenance. Instead, we used qualitative analysis to extract and select codes from  interviews~\cite{Saldana2011}. This allowed us to categorize conceptual elements (from coding) into findings. Also, categories emerged and evolved during analysis. We then represented our findings in an ``influence map'', complemented by a textual description.

\section{Findings}
\label{sec:map}

\begin{figure}[htb!]
    \centering
    \includegraphics[width=\linewidth]{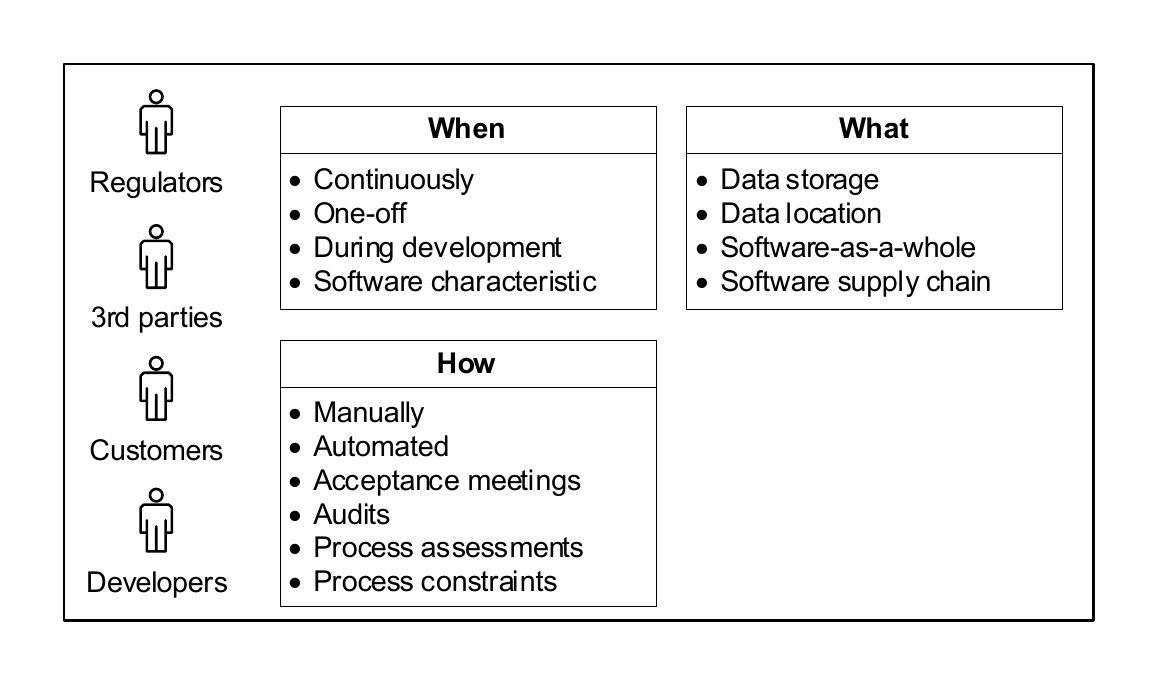}
    \caption{Provenance influence map}
    \label{fig:map}
\end{figure}

The resulting influence map (Figure~\ref{fig:map}) 
captures dimensions of provenance. 

\subsection{\emph{Who} Dimension}

This dimension captures who demands provenance for software. Provenance is mostly demanded from regulators (e.g., in banking) or other third parties. End users and developers rarely demand provenance information. 

We also identified provenance ``drivers'' which explain why organizations support provenance demanded by different stakeholders:
\begin{enumerate}
    \item Export restrictions:  Companies export products. Differences between countries can be of legal nature or in cultural requirements (e.g., language). For example, Australia sometimes has stricter requirements for banking software than the UK. Furthermore, authorities in countries have specific security requirements (e.g., the Australian Tax Office) or demand operational frameworks  (e.g., to support multi-factor authentication). Local regulations  (e.g., GDPR, SOC2, FDA regulations, or quality models such as ISO/IEC 27110 for cybersecurity) are dominating drivers for provenance. 
    \item Internal regulations: Other issues relate to regulations that are either internal to companies or the ``home market'' of companies. One example is the compatibility of licenses of components to build software systems. 

\end{enumerate}

Overall, compliance pressure is increasing and became a market entry barrier for new software providers. 

\subsection{\emph{When} Dimension}
This dimension describes when provenance is considered during the life time of software. It also describes when provenance impacts the development of software. We found that provenance might be required continuously rather than once as a ``one-off'' information. 

In detail, provenance impacts software (1) during development (i.e., it impacts how developers work, but is not necessarily relevant for users, e.g., requirements on data used during development or the development process followed), and (2) as characteristics of the final software product at runtime (e.g., the ability of software to explain why it behaved in a certain way).

These insights  influence techniques to assist provenance. For example, static techniques are applicable during development whereas dynamic techniques (such as application monitoring) are suitable for deployed and running systems.

\subsection{\emph{What} Dimension}
This dimension captures what is impacted by provenance in a software. We found that provenance is mostly about data storage technologies and storage location, and  security of software (but not algorithms). Furthermore, it can impact the whole software supply chain~\footnote{ Components, libraries, tools, and processes to develop, build, and publish software.}, including external components. As a consequence, provenance impacts generic quality attributes, such as security, data sovereignty and privacy. Data sovereignty not only impacts end users, but also developers during development, e.g., developers should not see client data when fixing bugs.

\subsection{\emph{How} Dimension}
This dimension describes how provenance information is managed (i.e., collected and analyzed). Provenance information can be collected manually or automated. We found few examples of automation or even tools (beyond SonarCube or internal tools to inspect code). Furthermore, provenance data is mostly captured in informal documents following company (not industry) standards. Companies do not expect a formal approach and create reports with consultants. This comes at a cost. 

We found the following practices to manage provenance information:
\begin{itemize}
\item Acceptance meetings: Companies conduct ``acceptance meetings'' with customers (e.g., involving an internal audit team that checks how software implements processes and handles data). 
\item Security audits: Security audits involve ``vulnerability management teams'' and focus on security of software. 
\item Process assessments: This involves control and testing teams that check that software engineering processes are robust by assessing ``internal audit points'', e.g., can one person trigger a change in the software that propagates to production; do architecture changes alert architecture team (e.g., modified dependencies). 
\item Process constraints: Organizations impose requirements on the development process as well as (reused) components. For example, a company in the banking sector reported that it uses different data for development than during operations (e.g., developers can only use bug reports if data are ``de-personalized''). Similarly, APIs are only available to certified ``digital service providers''. 
 Some companies define specific concepts (e.g., ``software of unknown provenance'').

\end{itemize}

Overall, companies consider trustworthiness as an overhead. It is perceived as another constraint that requires resources rather than a unique selling point. Automation might be useful, but is not a short-term objective. 

\subsection{A Note on Tools}

Tools can support all of the above discussed dimensions. However, there is no broad use of tools such as snyke or Dependabot (these could prevent  attacks like the Equifax incident~\footnote{https://www.doncio.navy.mil/Chips/ArticleDetails.aspx?ID=9457 [last access: September 6, 2022]}). As noted above, there is no significant uptake of ``advanced'' tools beyond standard tools, mostly due to a low (perceived) cost-benefit ratio. 
On the other hand, as mentioned above, we found some standard tools, e.g., SonarCube
, 
BlackDuck (to identify  used libraries), Veracode (for security scanning
) or Resharper 
to help refactor.

\section{Discussion and Conclusions}

We suggest that more guidance is needed to support provenance in software. Below we summarize take-aways:
\begin{itemize}
    \item 
    Provenance is demanded by regulators, not end users. 
    This is interesting given the studies on explainability as a software quality.
    \item 
    Provenance 
    becomes a differentiating factor for entry into new markets. However, it is currently considered as a burden rather than a unique selling point.
    \item Our third conclusion relates to challenges and inefficiencies related to dealing with provenance. Automation and the use of tools are not common.
\end{itemize}

In this paper we elicited the understanding of provenance needs from professionals in leadership roles. We did not get insights from developers involved in the details of implementing software. The focus on leaders is also the reason for the size of our sample (24 interviews). However, all participants were from different organizations. Also, we do not provide a detailed comparison with the literature. This is because we present a snapshot of the current state (and of perceptions) in the industry. 

In future work, we intend to develop techniques to ``retrofit'' provenance into existing software. We aim at retrofitting to reduce the effort of redesigning systems. Our emphasis will be on ensuring that the one-off effort to retrofit provenance is reasonable and that the runtime overhead and complexity do not increase too much. Our initial efforts utilized ideas from static program analysis.

\section*{Acknowledgments}
This research was  funded by the New Zealand Ministry of Business, Innovation \& Employment via the SfTI Veracity Technology Spearhead.

\bibliographystyle{ieeetr}
\bibliography{references}


\end{document}